\documentclass[aps,pre,reprint,groupedaddress,longbibliography,floatfix]{revtex4-1}

\usepackage{amsmath}
\usepackage{graphicx,psfrag}
\usepackage[caption=false]{subfig}
\usepackage{verbatim}

\begin{document}

%\preprint{}

\title{Integer Lattice Gas with Monte Carlo collision operator recovers the entropic lattice Boltzmann method with Poisson distributed fluctuations}


\author{Thomas Blommel}
\email{thomas.blommel@ndsu.edu}
\author{Alexander J. Wagner}
\email[]{alexander.wagner@ndsu.edu}
\homepage[]{www.ndsu.edu/pubweb/$\sim$carswagn}
%\thanks{}
\affiliation{Department of Physics, North Dakota State University, Fargo, North Dakota 58108, USA}


\date{\today}

\begin{abstract}
We are examining a new kind of lattice gas that closely resembles modern lattice Boltzmann methods. This new kind of lattice gas, that we call a Monte Carlo Lattice Gas, has interesting properties that shed light on the origin of the multi-relaxation time collision operator and it derives the equilibrium distribution for entropic lattice Boltzmann. Furthermore these lattice gas methods have Galilean invariant fluctuations given by a Poisson statistics, giving further insight into the properties that we should expect for fluctuating lattice Boltzmann methods. 
\end{abstract}

%\pacs{}

\maketitle

\section{Introduction}
The origin of fluctuations can be traced back to the discrete nature of matter\cite{einstein1906theorie}. When one examines a small enough systems, these fluctuations, originating from the stochastic nature of the dynamics of the discrete particles, manifest themselves. With few exceptions, \textit{e.g.} the dynamics close to a critical point\cite{kawasaki1970kinetic}, these fluctuations become irrelevant for larger systems, and the dynamics of this large-scale system becomes deterministic, allowing us to describe the system as a continuum. These continuum equations, like the continuity and Navier-Stokes equations for fluids, are enormously successful at describing macroscopic phenomena. So it is tempting to continue to utilize these continuous equations of motions down to scales where fluctuations become important. In these cases fluctuations, that were eliminated, have to be somehow re-introduced. One way to consistently introduce fluctuations is the Langevin approach\cite{langevin}, which consists of introducing a fluctuating term in the equations of motion and then adjusting the amplitude of the fluctuating term to give correct fluctuations in equilibrium. This is easily done since it is known that the states of the system obey a Boltzmann statistics.

This general narrative is the template for the development of fluctuating lattice Boltzmann methods. The original lattice gases introduced by Frisch, Hasslacher and Pomeau \cite{frisch1987lattice} consisted of a hexagonal lattice and particles moving along the links between nearest neighbor sites. The model was restricted to having at most one particle moving along each link. While it could be shown that this model had enough symmetry to recover the Navier-Stokes equations in the hydrodynamic limit, the model contained very large fluctuations, that needed to be averaged out to examine hydrodynamic phenomena. Part of the derivation of the Navier-Stokes equations required taking a formal ensemble average of the lattice gas, leading to a Boltzmann equation for the lattice gas\cite{frisch1987lattice}. Mc Namara \cite{mcnamara1988use} then realized that one could use this Boltzmann averaged lattice gas as a method in its own right. This method became known as the lattice Boltzmann method. The original lattice Boltzmann method was an exact ensemble average of the lattice gas, retaining the unconditional stability of the lattice gas as well as several of its flaws. At the time a key advantage of the new lattice Boltzmann method was the elimination of all noise.

Soon it was discovered that the lattice Boltzmann method could be simplified by using a Bhatnagar-Gross-Krook (BGK) approximation for the collision operator \cite{Higuera1989}. The idea here is that the distribution function will approach the local equilibrium distribution, consistent with the locally conserved density and momentum. This significantly simplified the collision operation and at the same time removed some undesirable artifacts that had survived the transition from the lattice gas to the lattice Boltzmann methods, like the velocity dependence of the viscosity\cite{frisch1987lattice}. While the improvements of these novel lattice Boltzmann methods made them extremely popular any application that needed fluctuations was stuck with lattice gas methods. Tony Ladd had utilized the fluctuations of lattice gases to simulate the Brownian motion of colloids \cite{ladd1988application}. He then developed a fluctuating version of the lattice Boltzmann method by including a fluctuating stress tensor that would recover the fluctuating hydrodynamic equations \cite{ladd1993short}. While this recovered the correct hydrodynamic limit, the fluctuations could be seen to be only correct for wavelength of the size of the system. Fluctuations at smaller scales were suppressed. The reason for this became clear when Adhikari \textit{et al.} noticed that all kinetic modes, not only those related to the hydrodynamic quantities, needed to have fluctuations added\cite{adhikari2005fluctuating}. The theoretical framework for this was work on generating a fluctuating version of the linearized Boltzmann equation by Bixon and Zwanzig\cite{bixon1969boltzmann}.  Much work has gone into examining the correct form of these flucuations for lattice Boltzmann since then\cite{duenweg2007statistical,kaehler2013fluctuating,wagner2016fluctuating}.
The remaining difficulties with defining consistent fluctuations for lattice Boltzmann systems also hinder the progress for extended version of the fluctuating lattice Boltzmann method for non-ideal systems \cite{gross2010thermal, ollila2011fluctuating,thampi2011lattice,gross2012simulation,belardinelli2015fluctuating}.

One obvious extension to lattice gases that maintains their discrete properties, is to extend the occupation numbers from booleans to integers. In 1988, just one year after the FHP paper \cite{frisch1987lattice}, Kim Molvig \textit{et al.} pioneered lattice gases that were able to remove lattice artifacts in the pressure and viscosity \cite{Molvig88}, although the results are more sketched than presented in this article. This article was later credited by Chen \textit{et al.} to have introduced an approach called ``Digital Physics''. This approach was the idea behind the founding of the EXA company in 1991. There are some indications of further developments of this approach, notably the transition from boolean to integer occupation numbers, in subsequent papers \cite{chen1997digital,teixeira1997digital,anagnost1997digital,halliday1999simulation,succi1999integer} but most of the details of the approach remain unpublished since they formed the basis for proprietory technology. The paper by Chen \cite{chen1997digital} sketches an approach of extending boolean lattice gases to lattice gases with integer occupation numbers reminiscent of the one presented here, although it appears that it concerns an algorithm that appart from mass and momentum also conserves energy, rather than the approach presented here that, like most standard LB approaches, replaces conservation of energy with an isothermal approach. We will comment further on the similarities and differences later in the paper.

The integer lattice gas approach was independently re-introduced by Boghosian \textit{et al.} in 1997 \cite{boghosian1997integer}. The focus of this paper is the thermodynamics behavior of integer lattice gases with the finite number of bits. While the theoretical part focuses on energy conserving systems, the explanation of the sampling method for the collision considers only mass and momentum conservation. The collision process then consists of a non-trivial sampling of points from a polyotope of allowed states, the details of which remain unpublished.

One year later Chopard and Masselot \cite{chopard1998multiparticle,masselot1998multiparticle} describe a different integer lattice gas approach: they impose a collision operator that closely mimicks a standard lattice Boltzmann BGK collision operator with an equilibrium distribution that is quadratic in the velocities. The result of this LB inspired collision process is then a continuous probability distribution that is sampled to obtain a new discrete distribution. The sampling process, however, will lead to a violation of momentum conservation, so an additional random walk process is required until a distribution is found that is consistent with the original momentum. This method shows a surprising density dependence of the viscosity \cite{masselot1998multiparticle} that the authors attribute to fluctuation effects. 

Because of the rise of lattice Boltzmann, however, the development of lattice gases was then significantly curtailed. Even the EXA company switched to lattice Boltzmann implementations, and little has been done on the development of integer lattice gases since these early days. During his Diplomarbeit, Marin Geier worked on integer equivalents of the lattice Boltzmann approach, but he judged this early efforts as mostly unsuccessful. Some remnants of the integer representation can be seen in \cite{geier2006cascaded}, but here the appearance of integer rather than continuous distributions is incidental rather than central to the approach and was given up in further developments of the method in favor of continuous distribution functions.

Our renewed interest in integer lattice gas methods stems from our interest in fluctuations. We wanted to get away from the approach of introducing fluctuations in a continuous system through a Langevin approach, and back to a more natural approach of obtaining fluctuations as a direct result of the discrete nature of the fluid we are examining. The idea of using a lattice gas as an starting point to derive fluctuating lattice Boltzmann methods was pioneered by Duenweg \cite{duenweg2007statistical} who proposed a lattice gas with integer (rather than boolean) occupation numbers and Monte Carlo collision operator based on a zero-velocity equilibrium distribution as a starting point to derive fluctuating lattice Boltzmann methods. But this approach remained entirely theoretical without any attempt at an actual implementation, similar to the original introduction of lattice Boltzmann as a theoretical tool to derive the hydrodynamic limit of lattice gas methods.

The aim of this paper is to derive a discrete lattice gas version that corresponds to current state of the art lattice Boltzmann methods and which would have (at least in close approximation) these lattice Boltzmann methods as its ensemble average and its integer implementation shows Galilean invariant fluctuations of an ideal gas. This means the occupation numbers should be Poisson distributed in equilibrium. The results are very encouraging: we show below that the Monte Carlo Lattice Gas introduced in this paper recovers the equilibrium distribution of the entropic lattice Boltzmann method \cite{ansumali2003minimal,Chikatamarla2006} as its ensemble average, and recovers Galilean invariant fluctuations at even better accuracy than the approach by Kaehler \textit{et al.} \cite{kaehler2013fluctuating}. The collision operator is to first order a multi-relaxation time BGK approach, but it has additional second order terms.

This paper is structured as follows: we first introduce the basic lattice Boltzmann method and then use these result to propose a lattice gas method with a Monte Carlo collision operator. We then show how this is practically implemented in one and two dimensions. The Boltzmann limit of this lattice gas implementation is derived and we show that it recovers the entropic lattice Boltzmann method (rather than a lattice Boltzmann method with a quadratic equilibrium that inspired it) in the limit of small deviations from local equilibrium. The last section shows that this implementation indeed recovers the independent Poisson statistics for the densities exactly; this is the statistics of an ideal gas that we previously tried to impose on lattice Boltzmann methods, but only recovered approximately \cite{kaehler2013fluctuating}. 

\section{Basic lattice Boltzmann}
The lattice Boltzmann algorithm consists of continuous densities $f_i$ that are associated with velocities $v_i$ which are lattice velocities. This means that if $x$ is a lattice position, $x+v_i$ will also be a lattice position. These densities are defined on each lattice point $x$ at discrete times $t$ which are taken to be integer values. They evolve in time according to the lattice Boltzmann equation
\begin{equation}
  f_i(x+v_i, t+1) = f_i(x,t)+\Omega_i
  \label{LB}
\end{equation}
where $\Omega_i$ is the collision operator. In its multi-relaxation-time (MRT) BGK form this collision operator can be written as
\begin{equation}
  \Omega_i = \sum_j \Lambda_{ij} [f_i^0-f_i(x,t)]+\xi_i
\end{equation}
where $f_i^0$ is the local equilibrium distribution and $\xi_i$ is a noise term. Most lattice Boltzmann methods are conserving mass and momentum, but instead of conserving energy they are coupled to a heat bath. We define the local density $\rho$ and momentum density $\rho u$ through the velocity moments of the $f_i$ as 
\begin{align}
  \rho &= \sum_i f_i,\\
  \rho u_\alpha &= \sum_i f_i v_{i\alpha}.
\end{align}
Mass and momentum conservation is ensured by the requirement that the local equilibrium distribution also obey
\begin{align}
  \rho &= \sum_i f_i^0,\\
  \rho u_\alpha   &= \sum_i f_i^0 v_{i\alpha}.
\end{align}
For the recovery of the Navier-Stokes equations we also require higher order moments for the equilibrium distribution. From analogy to the velocity moments of the Maxwell-Boltzmann distribution we demand
\begin{align}
  \sum_i f_i^0 (v_{i\alpha}-u_\alpha)(v_{i\beta}-u_\beta) &= \rho \theta \delta_{\alpha\beta}
  \label{eqn:f0_2mom}\\
  \sum_i f_i^0 (v_{i\alpha}-u_\alpha)(v_{i\beta}-u_\beta)(v_{i\gamma}-u_\gamma) &= Q_{\alpha\beta\gamma}  
\end{align}
where $Q_{\alpha\beta\gamma}$ should be zero. Unfortunately for small velocity sets typically used which have $v_{i\alpha}^3=v_{i\alpha}$ because $v_{i\alpha}\in\{-1,0,1\}$ this is not possible \cite{wagner2006investigation}. But for the special choice of $\theta=1/3$ it can be reduced to $\rho u_\alpha u_\beta u_\gamma$, which is assumed to be small for $u\ll 1$ found in lattice Boltzmann simulations.
An expansion of the Maxwell Boltzmann distribution to second order in velocities then gives
\begin{equation}
  f_i^0(\rho,u)=\rho w_i\left(1+\frac{v_{i\alpha}u_\alpha}{\theta}+\frac{v_{i\alpha}u_\alpha v_{i\beta}u_\beta}{2\theta^2}-\frac{u_\alpha u_\alpha}{2\theta}\right).
  \label{LBequil}
\end{equation}
The $w_i$ are obtained by matching the discrete velocity moments to the continuous velocity moments of the equilibrium distribution. To recover the continuity and Navier-Stokes equations we require the first four moments. All odd moments are zero and the even moments obey
\begin{align}
  \sum_i w_i &= 1\\
  \sum_i w_i v_{i\alpha}v_{i\beta} &= \theta \delta_{\alpha\beta}\\
  \sum_i w_i v_{i\alpha}v_{i\beta}v_{i\gamma}v_{i\delta} &= \theta^2 (\delta_{\alpha\beta}\delta_{\gamma\delta}+\delta_{\alpha\gamma}\delta_{\beta\delta}+\delta_{\alpha\delta}\delta_{\beta\gamma})
\end{align}
The collision matrix $\Lambda_{ij}$ is constructed so that the stress moments are relaxed at a rate that determines the viscosity, and other moments can be relaxed at different rates \cite{d1994generalized,lallemand2000theory}, \textit{e.g.} to optimize the stability of the method. In general we can write the moments as
\begin{equation}
  M^a = \sum_i m_i^a f_i
\end{equation}
where the first moments will be related to the hydrodynamic moments. For completeness we require as many moments $M^a$ as we have densities $f_i$ so we can have a one-to-one correspondence between moment and velocity space. It is often useful to require that the square matrix generating the moments be orthogonal with respect to some measure. Particularly for fluctuating applications it is often \cite{adhikari2005fluctuating,kaehler2013fluctuating,kaehler2013derivation} found to be advantageous to require
\begin{equation}
  \sum_a w_i m_i^a m_j^a = \delta_{ij}.
  \label{eqn:scalar}
\end{equation}
This implies the backtransform
\begin{equation}
  f_i = \sum_a w_i m_i^a M^a
\end{equation}
as well as
\begin{equation}
  \sum_i w_i m_i^a m_i^b = \delta^{ab}.
\end{equation}
This implies that moments $m_i^a$ are constructed startig from the hydrodyanmic moments of interest (density, momentum, stress tensor), complemented by a set of ghost-modes, which are then orthonormalized with a Gram-Schmidt orthonomalization scheme using the scalar product implied by (\ref{eqn:scalar}).
In this representation the collision matrix is designed to be diagonal:
\begin{equation}
  \Lambda^{ab}=\sum_{i,j} m_i^a\Lambda_{ij}w_j m_j^b = \frac{1}{\tau^a} \delta^{ab}.
\label{eqn:Lambda}
\end{equation}
Setting the $\tau^a$ then fully determines the algorithm.
Next we will briefly discuss two common lattice Boltzmann velocity sets for one and two dimensions.

\subsection{D1Q3}
The minimal lattice lattice Boltzmann method in one dimension consists of only three velocities corresponding to particle moving to a lattice site to the right, the lattice site to the left or a particle remaining at its lattice site. This is referred to as the D1Q3 model. Our velocity set is given by
\begin{equation}
  v_{ix}\in\{-1,0,1\}.
\label{D1Q3vel}
\end{equation}
The weights are given by
\begin{align}
  w_0&=2/3 \nonumber\\
  w_{-1}=w_1&=1/6
  \label{D1Q3wi}
\end{align}
The moment matrix is given by
\begin{equation}
  M = \left(
  \begin{array}{rrr}
    1& 1& 1\\
    -\sqrt{3}& 0 & \sqrt{3}\\
    \sqrt{2} & -\sqrt{\frac{1}{2}} & \sqrt{2}
  \end{array} \right)
  \label{mD1Q3}
\end{equation}
The first two rows correspond to the conserved mass and momentum moments respectively. So there is only one relaxation time for the moment related to the third row for the D1Q3 model.

\begin{figure}
  \centering
  \includegraphics[width=0.25\columnwidth]{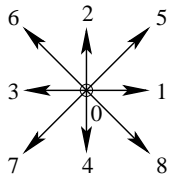}
  \caption{Numbering scheme for the D2Q9 velocity set.}
  \label{fig:D2Q9vel}
\end{figure}

\subsection{D2Q9}
In the D2Q9 implementation, our velocity set consists of 9 velocities given by the product set of the D1Q3 velocities of Eq. (\ref{D1Q3vel}):
\begin{equation}
  v_i=
  \begin{bmatrix}
    v_x\\
    v_y
  \end{bmatrix}
  \label{D2Q9vi}
\end{equation}
where $v_x,v_y\in\{-1,0,1\}$.
The weights in the two-dimensional case are just products of the one-dimensional weights
\begin{align}
  w_{(v_x,v_y)} = w_{v_x} w_{v_y}
\end{align}
or, using the numbering of Fig.~\ref{fig:D2Q9vel}, the more standard
\begin{align}
  w_0 &= 4/9,\\
  w_{1-4} &= 1/9,\\
  w_{5-8} &= 1/36.
\end{align}
The moment matrix is given by
\begin{equation}
  M = \left(\begin{array}{ccccccccc}
    1  & 1  & 1  & 1 & 1 & 1 & 1 & 1 & 1\\
    0  & \sqrt{3}&0&-\sqrt{3}&0&\sqrt{3}&-\sqrt{3}&-\sqrt{3}&\sqrt{3}\\
    0&0&\sqrt{3}&0&-\sqrt{3}&\sqrt{3}&\sqrt{3}&-\sqrt{3}&-\sqrt{3}\\
    0&\frac{3}{2}&-\frac{3}{2}&\frac{3}{2}&-\frac{3}{2}&0&0&0&0\\
    0&0&0&0&0&3&-3&3&-3\\
    -1&\frac{1}{2}&\frac{1}{2}&\frac{1}{2}&\frac{1}{2}&2&2&2&2\\
    0&-\sqrt{\frac{3}{2}}&0&\sqrt{\frac{3}{2}}&0&\sqrt{6}&-\sqrt{6}&-\sqrt{6}&\sqrt{6}\\
    0&0&-\sqrt{\frac{3}{2}}&0&\sqrt{\frac{3}{2}}&\sqrt{6}&\sqrt{6}&-\sqrt{6}&-\sqrt{6}\\
    \frac{1}{2}&-1&-1&-1&-1&2&2&2&2
  \end{array}\right)
  \label{MomD2Q9}
\end{equation}
where the numbering of the rows and columns is given by the numbering of the velocities in Figure \ref{fig:D2Q9vel}.
We then get the moments
\begin{equation}
  M^a = \sum_i m_i^a f_i = \left(\begin{array}{c}
    \rho\\
    j_x\\
    j_y\\
    \sigma_-\\
    \sigma_{xy}\\
    \sigma_+\\
    q_x\\
    q_y\\
    t
  \end{array}\right)
  \label{eqn:D2Q9moments}
\end{equation}

This concludes our very brief recap of the lattice Boltzmann algorithm (in its most common form) and we will next present the Monte Carlo lattice Gas algorithm introduced in this paper.

\section{The Monte Carlo Lattice Gas algorithm}
The main idea behind defining a lattice gas that behaves equivalently to a lattice Boltzmann method is that we will use a Monte Carlo collision operator that explicitly conserves mass and momentum and will recover the correct equilibrium distribution for $u=0$. At each lattice site $x$ the number of particles that streamed in from a lattice position $x-v_i$ is denoted as $n_i(x,t)$. This number is an integer, in contrast to the real number $f_i(x,t)$ in the lattice Boltzmann approach. These particles then are re-distributed due to collisions, and this re-distribution will be denoted as $\Xi_i$. We then have the evolution equation
We write the evolution equation as
\begin{equation}
  n_i(x+v_i, t+1) = n_i(x,t)+\Xi_i
\end{equation}
which looks equivalent to the lattice Boltzmann equation (\ref{LB}), except that the occupation numbers $n_i$ are integers, and that the collision term is an inherently probabilistic term, not a deterministic one as in the lattice Boltzmann approach. We consider that this collision term is the accumulated effect of many two-particle collisions. This may not be the most numerically efficient algorithm, but it avoids complexities encountered by Boghosian \cite{boghosian1997integer} and Duenweg \cite{duenweg:2016} because it is conceptually clean. It may be similar to the approach of Digital Physics, as hinted at in \cite{chen1997digital}, but we have been unable to uncover the details of this approach.

To define the collision term $\Xi_i$ we demand that it recover the zero-velocity equilibrium distribution of lattice Boltzmann. From Eq. (\ref{LBequil}) we see that it is given simply by
\begin{equation}
f_i^{eq}(\rho,u\equiv0) = \rho w_i.
\end{equation}
Let us now consider a collision of two particles with velocities $v_i$ and $v_j$. 
We denote the probability of colliding these two particles and ending up with two particles with velocities $v_k$ and $v_l$ with $P_{ij\rightarrow kl}$. We assume detailed balance, which means that the forward and backward collisions times the probabilities of finding these pairs in equilibrium have to be equal. Let us a assume that the number of particles at a lattice site is $N$. At one lattice site the probability of picking a particle with velocity $v_i$ is then $w_i$.

We can then write the detailed-balance condition for the equal probability for the forward and backward collisions as
\begin{equation}
 w_i w_j P_{ij\rightarrow kl} = w_k w_l P_{kl\rightarrow ij}.
\end{equation}
This fixes the ratios of the forward and backward collisions to
\begin{equation}
  \frac{P_{ij\rightarrow kl}}{P_{kl\rightarrow ij}} = \frac{w_k w_l}{w_iw_j}.
  \label{transprob}
\end{equation}
Assuming that we do not have any additional conserved quantities this condition will ensure that the system will approach the lattice Boltzmann equilibrium distribution if the system has a  mean velocity of zero.
For actual collisions, \textit{i.e.} $ij\neq kl$, we pick the transition probability
\begin{equation}
  P_{ij\rightarrow kl}\propto \min\left(1,\frac{w_kw_l}{w_iw_j}\right) \delta_{(v_i+v_j),(v_k+v_l)}
  \label{collcons}
\end{equation}
where $\delta$ is the Kronecker delta. Note that this ensures the probability ratio requirement of Eq. (\ref{transprob}), which is always guaranteed for the case $ij=kl$. Next we have to ensure that the probabilities add up to one:
\begin{equation}
  \sum_{kl} P_{ij\rightarrow kl} = 1
\end{equation}
and of course we also require $P_{ij\rightarrow kl}\geq 0$. Formally we can achieve this by introducing a proportionality factor $\lambda_{ij,kl}$ to get
\begin{equation}
  P_{ij\rightarrow kl}= \left\{ \begin{array}{cc}
    \lambda_{ij,kl} \min\left(1,\frac{w_kw_l}{w_iw_j}\right) \delta_{(v_i+v_j),(v_k+v_l)}& ij\neq kl\\
    1-\sum_{k'l'\neq ij}P_{ij\rightarrow k'l'} & ij=kl
    \end{array}\right.
\end{equation}
and requiring $\lambda_{ij,kl}=\lambda_{kl,ij}$ to ensure Eq. (\ref{transprob}). The $\delta$-function ensures that only collisions that conserve momentum (mass conservation is trivially true) have a non-zero probability. Thus the collision will conserve mass and momentum.
Note that the lower condition, together with the requirement of positive probabilities,  puts an constraint for the largest achievable $\lambda_{ij,kl}$.

The lattice gas algorithm then consists of moving particles according to their associated velocities $v_i$ and then picking $C$ pairs of particles at random and colliding them according to the transition probabilities given above. In order to maintain conceptual clarity we use the naive, straight forward, non-optimized algorithm. As explained above, we denote the number of particles at position $x$ and time $t$ moving in direction $v_i$ with $n_i(x,t)$. The local density is then given by
\begin{equation}
  N(x,t) = \sum_i n_i(x,t).
\end{equation}
At each lattice site we then pick a pair of particles to collide by selecting two evenly distributed random numbers between 1 and $N$. Given these two random numbers $r_1$ and $r_2$ we identify the identity of these particles by
\begin{equation}
  r_1\rightarrow s_1 \mbox{ for }\sum_{i=0}^{s_1-1}n_i < r_1 <\sum_{i=0}^{s_1} n_i.
  \label{eqn:select}
\end{equation}
This then implies two different random numbers $s_1$ and $s_2$, corresponding to the number of the two selected velocities. 
The two random numbers then imply picking two particles with velocities $v_{s_1}$ and $v_{s_2}$. For a collision of this pair of particles we then pick two outgoing velocities $v_{s_3}$ and $v_{s_4}$ with probability $n_{s_1} n_{s_2} P_{s_1 s_2\rightarrow s_3 s_4}/N^2$. The effect of the collision number $c$ is then given by a random variable $\vartheta_i(s_1,s_2,s_3,s_4)$ that takes on the value
\begin{equation}
  \vartheta_i(s_1,s_2,s_3,s_4) = (\delta_{is_3}+\delta_{is_4}-\delta_{is_1}-\delta_{is_2}).
\label{omega}
\end{equation}
which corresponds to the change in the number of particles corresponding to velocity $v_i$ after the collision $(v_{s_1}v_{s_2}\rightarrow v_{s_3} v_{s_4})$ has been chosen.
This process is repeated $C$ times, and we get the full collision operator as the sum of these random variables:
\begin{equation}
\Xi_i = \sum_{c=1}^C \vartheta_i(s_1,s_2,s_3,s_4).
\end{equation}
where it is understood that each of these sub-collision operators $\vartheta_i(s_1,s_2,s_3,s_4)$ is the result of a collision for the state given by the previous sub-collisions, according to Eq. (\ref{eqn:select}). Note that this collision operator explicitly conserves mass and momentum because of Eq. (\ref{collcons}).

Next we will show two explicit implementations of this algorithm in one and two dimensions for standard velocity sets.

\subsection{A D1Q3 implementation}
A state is given by the numbers of particles $n_i$ moving in direction $v_i$. 
For this velocity set the only collision that conserves momentum and changes the state of the system consists of two particles moving in opposite directions that come to rest, or the inverse process of two rest particles that will move appart in opposite directions. 
\begin{align}
  \{-1,1\}&\rightarrow\{0,0\}\\
  \{0,0\}&\rightarrow\{-1,1\}
\end{align}
For the D1Q3 lattice Boltzmann method the weights are given by Eq. (\ref{D1Q3wi}). We get for the transition probabilities
\begin{align}
  P_{-11\rightarrow 00} &=\lambda\\
  P_{-11\rightarrow -11}&=1-\lambda\\
  P_{-11\rightarrow 1-1}&=0\\
  P_{1-1\rightarrow 00} &=\lambda\\
  P_{1-1\rightarrow 1-1}&=1-\lambda\\
  P_{1-1\rightarrow -11}&=0\\
  P_{00\rightarrow 00} &= 1-\lambda/8\\
  P_{00\rightarrow -11} &=\lambda/16\\
  P_{00\rightarrow 1-1} &=\lambda/16
\end{align}
where we have set some irrelevant switching probabilities to zero since they do not change the state. 

\begin{figure}
  \centering
  \includegraphics[width=0.9\columnwidth]{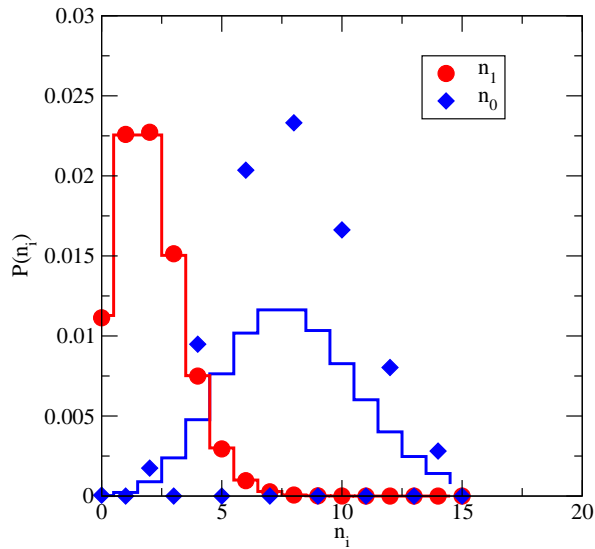}
  \caption{Poisson-like distributions with additional constraint are found for the D1Q3 model. Symbols represent simulation results, and solid lined represent the Poisson distribution corresponding to the expected equilibrium distribution.  All lattice sites were initialized with even number of rest particles, and this property is conserved by collisions.  Here $P(n_1)=P(n_{-1})$.  Note that the difference between the actual distribution and the Poisson distribution does not affect the distribution of moving particles.}
  \label{fig:1dPoisson}
\end{figure}

As a consequence of this small collision set, an accidentally conserved quantity is introduced. The number of rest particles at each lattice site will stay either even or odd throughout the entire simulation, as they can only be created or destroyed in pairs.  This conserved quantity prevents us from exactly recovering the equilibrium fluctuations, particularly the Poisson distributions. This is shown in Fig. \ref{fig:1dPoisson} for a worst-case scenario where all lattice sites have an even number of rest-particles. For this simulation, we used $L_x=100$, $C=10$, $\rho=12$, allowed the simulation to equilibrate for 20000 time steps, and averaged over 80000 time steps.

\subsection{A D2Q9 implementation}
The velocities of Eq. (\ref{D2Q9vi}) allow for many more collisions, and these collisions can be grouped into equivalence classes, in which each collision within a class is simply a rotation of the other collisions. In order to ensure isotropy, we must ensure that the $\lambda_{ij,kl}$ for all collisions in an equivalence class are the same.

Some collisions in our D2Q9 model break the conservation of `even or odd-ness' of $n_0$ at each lattice site.  However, in the projection of our simulation along the x or y direction, we still have that each row and column is constant in it's 'even or odd-ness'.  This has a much smaller impact on the simulation as our lattice size grows. Let us assume the lattice has $L_x$ lattice sites in the x-direction and $L_y$ lattice sites in the y-direction.  Then the number of conserved quantities grows only as $L_x+L_y$, but the number of lattice sites grows as $L_x L_y$, so the fraction of spuriously conserved degrees of freedom becomes small for large lattices.

\begin{figure}
  \centering
  \includegraphics[width=0.9\columnwidth]{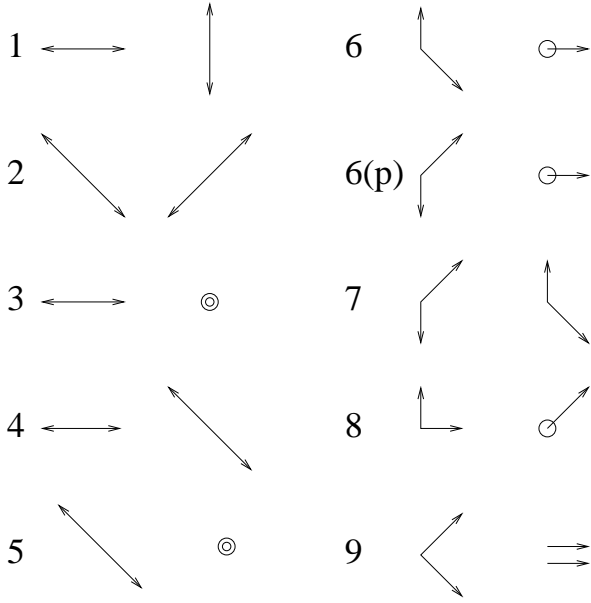}
  \caption{The 9 equivalence classes for binary collisions that both change the state and conserve momentum. Arrows correspond to particles moving with the corresponding velocity $v_i$, circles denote rest-particles. Each of these collision equivalence classes is associated with a different rate constant $\lambda_{ij,kl}$. The numbers are used to identify the relaxation rates $\lambda_i$ corresponding to each equivalence class.  Versions of these collisions rotated by multiples of $90^\circ$ are implied.  6 and 6(p) are related by parity and have the same relaxation rate. Note that 1--5 have a net momentum of zero, 6--7 have a total momentum of (1,0), 8 has a momentum of (1,1) and 9 corresponds to a total momentum of (2,0).}
  \label{fig:D2Q9coll}
\end{figure}

The equivalence classes are shown schematically in Fig. \ref{fig:D2Q9coll}. As a small side note, the the collision 9 was first proposed by Molvig \cite{Molvig88}. 
This allows us to write down all possible binary collisions and their probabilities in principle, exactly in the same way as we did for the D1Q3 model. We make the simplification that the relaxation rates $\lambda_{ij,kl}$ are instead referred to as $\lambda_i$, where each index $i$ refers to the number of the equivalence class identified in Fig. \ref{fig:D2Q9coll}. We now have many more collisions, which we will not explicitly list their probabilities here. It should also be noted that there are three and more particle collisions that could be included. For simplicity those are neglected here.

\section{Boltzmann average of the lattice gas}
To compare the lattice gas results to the lattice Boltzmann method we will now examine an non-equilibrium ensemble average of the lattice gas evolution equation. We define the particle probability densities as 
\begin{equation}
  f_i(x,t) = \langle n_i(x,t) \rangle
\end{equation}
where $\langle \rangle$ implies a non-equilibrium average over an ensemble of microscopic realizations leading to the same macroscopic state. We define the density as
\begin{equation}
\rho(x,t) = \langle N(x,t)\rangle.
\end{equation}
The evolution equation for the average particle densities is then
\begin{equation}
  f_i(x+v_i,t+1) = f_i(x,t)+\Omega_i.
  \label{MCLB}
\end{equation}
The collision operator is the averaged version of the lattice gas collision operator
\begin{equation}
  \Omega_i = \langle\; \Xi_i\rangle.
\end{equation}
We can define intermediate distribution functions recursively as distribution functions affected by one single collision
\begin{align}
  f_i^{(n+1)} &= f_i^{(n)}+\sum_{jklm}\vartheta_i(jklm) \frac{ f_j^{(n)} f_k^{(n)} }{\rho^2} P_{jk\rightarrow lm}\\
  &= f_i^{(n)}+\Omega^{(n)}_i
\label{OmN}
\end{align}
where it is understood that
\begin{equation}
  f^{(0)}_i = f_i(x,t).
\end{equation}
The averaged collision operator can then be written as
\begin{align}
  \Omega_i = \sum_{n=1}^C \Omega^{(n)}_i .
  \label{eqn:avcoll}
\end{align}
From this collision operator we can obtain the equilibrium distributions for the $f_i$ by demanding
\begin{equation}
  0 = \sum_{jklm} \vartheta_i(jklm) \frac{ f_j^{eq} f_k^{eq} }{\rho^2} P_{jk\rightarrow lm}.
\end{equation}
Note that it is sufficient for this purpose to consider a single sub-collision of Eq. (\ref{OmN}), rather than the full collision operator of Eq. (\ref{eqn:avcoll}).
This gives us a quadratic matrix equation in the equilibrium density. One might expect not to obtain a unique solution. However, since we require that $f_i^{eq}>0$ we will find below that only one physical solution survives.  

\subsection{D1Q3 results}
In the one-dimensional case of D1Q3 we find that Eq. (\ref{eqn:avcoll}) gives
\begin{equation}
\Omega^{(1)}(\{f_i\})=\frac{\lambda}{\rho^2}\begin{bmatrix}
 \frac{f_0^2}{8}-2{f_{-1}} {f_1} \\
 4 {f_{-1}} {f_1}-\frac{{f_0}^2}{4} \\
 \frac{{f_0}^2}{8}-2{f_{-1}} {f_1} \\
\end{bmatrix}.
\label{eqn:collmatD1Q3}
\end{equation}
The equilibrium distribution has to be found with respect to the corresponding conserved quantities of mass $N(x,t)$ and the momentum
\begin{equation}
  N(x,t) U(x,t) = \sum_i n_i(x,t) v_i.
\end{equation}
This corresponds to an ensemble averaged momentum of
\begin{equation}
  \rho(x,t) u(x,t) = \langle N(x,t)U(x,t)\rangle.
\end{equation}
Furthermore we define a second moment
\begin{equation}
  \pi =\sqrt{2}( f_{-1}-f_0/2+f_1).
\end{equation}
We can use a matrix that transforms the $f_i$ vector onto the conserved moments and one free moment of Eq. (\ref{mD1Q3}).

We then have the moments
\begin{equation}
  M
\begin{bmatrix}
 f_{-1}\\
 f_0\\
 f_1\\
\end{bmatrix}
=
\begin{bmatrix}
\rho\\
\sqrt{3}\rho u\\
\pi\\
\end{bmatrix}.
\end{equation}
We can then write the collision operator for a single collision of Eq. (\ref{OmN}) in this moment space as
\begin{align}
  M\Omega_i^{(1)}(f_i)
&= \frac{\lambda}{\rho^2}\begin{bmatrix}
0\\
0\\
\frac{3}{\sqrt{32}}\left(f_0^2-16f_{-1}f_1\right)
\end{bmatrix}\nonumber\\
&=\frac{\lambda}{\rho^2}
\begin{bmatrix}
0\\
0\\
\frac{6u^2\rho^2-\pi(\pi+\sqrt{8}\rho)}{\sqrt{8}}\\
\end{bmatrix}.
\label{eqn:1Ddecay}
\end{align}
In equilibrium the collision operator does not change the distributions. Reversely, we can find the equilibrium distribution by demanding that the collison operator be zero. Thus we obtain $\pi^{eq}$ in terms of the conserved quantities as 
\begin{equation}
\pi^{eq}=\sqrt{2}\rho(-1\pm\sqrt{1+3u^2}).
\end{equation}
The solution with the negative sign leads to negative densities, while the other solution stays positive.  We obtain the D1Q3 equilibrium distribution
\begin{equation}
  \left[
    \begin{array}{c}
      f_{-1}^{eq}\\
      f_0^{eq}\\
      f_1^{eq}
    \end{array}
    \right]
  =
  \begin{bmatrix}
\frac{\rho }{6}\left(-1-3u+2\sqrt{1+3u^2}\right)\\
\frac{2 \rho }{3}\left(2-\sqrt{1+3u^2}\right)\\
\frac{\rho }{6}\left(-1+3u+2\sqrt{1+3u^2}\right))
\end{bmatrix}
\end{equation}
which can be written as
\begin{align}
  &f_i^{eq}(\rho,u_x)\nonumber\\
  =& \rho w_{v_{ix}}\left[1+3 v_{ix}u_x+(3v_{ix}^2-1)(\sqrt{1+3u_x^2}-1)\right].
  \label{eqn:D1Q3equil}
\end{align}
where we introduced a weight $w_{v_{ix}}$, anticipating that the weights for higher dimensions can be written as products of one-dimensional weights.
For $u=0$ we recover $f_i^{eq}=\rho w_i$, which was our starting point in selecting the collision probabilities. Once the collision probabilities are fixed, they also imply equilibrium distributions for different conserved velocities. This equilibrium distribution, and its dependence on $u$, is therefore the logical consequence of selecting the collision probabilities corresponding to the imposed zero velocity equilibrium distribution function. 

\begin{figure}
  \centering
  \includegraphics[width=0.9\columnwidth]{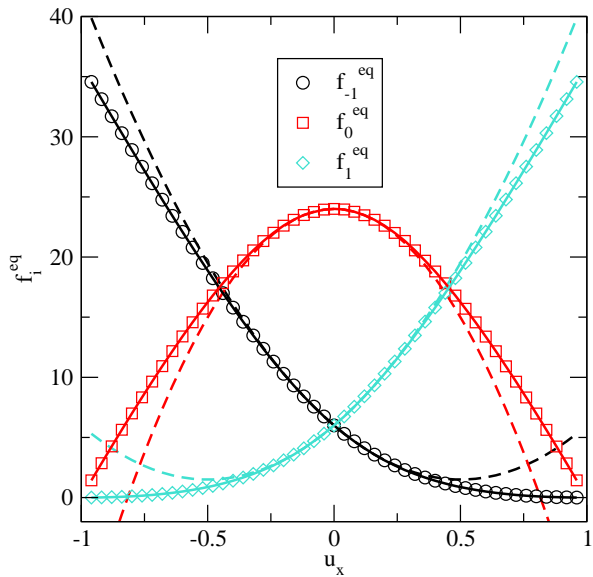}
  \caption{Comparison between the measured average $\langle n_i\rangle$ (shown as symbols) and the predicted equilibrium distribution, $f_i^{eq}$, of Eq. (\ref{eqn:D1Q3equil}) (shown as solid lines) for the D1Q3 model. This is also compared to the lattice Boltzmann equilibrium distribution, $f_i^0$, of Eq. (\ref{LBequil}) (dashed lines).}
  \label{fig:D1Q3equil}
\end{figure}

We show a comparison of a measured equilibrium distribution to this prediction in Fig. \ref{fig:D1Q3equil}. For the measured equilibrium distribution we show results for $\langle N \rangle=36$, $\lambda =1$, a lattice size of $L_x=100$ with periodic boundary conditions. We used $C=10$ collisions per iteration step and the simulation was initially run for $15,000$ steps to equilibrate the simulation, and then another $15,000$ iterations were used for the measurements. We find excellent agreement between this prediction and our measurements of the equilibrium distribution. The agreement between our new equilibrium distribution and that of a standard LB equilibrium distribution function of Eq. (\ref{LBequil}) is also close for $|u|<0.4$, but diverges thereafter.

This difference implies that the second moment of the equilibrium distribution will not obey the lattice Boltzmann requirement of Eq. (\ref{eqn:f0_2mom}) for all $u$. This is not surprising since the MCLG equilibrium distribution, unlike it's LB equivalent, is positive definite. This means that for a velocity of $u=\pm1$ we require $f_{\pm1}=\rho$, and therefore the second moment must be zero for this extreme value. However lattice Boltzmann simulations typically require $u\ll1$, which is usually taken as $u<0.1$, so disagreements between these equilibrium distributions outside the range $|u|<0.1$ have little practical relevance.

\begin{figure}
  \centering
  \includegraphics[width=0.9\columnwidth]{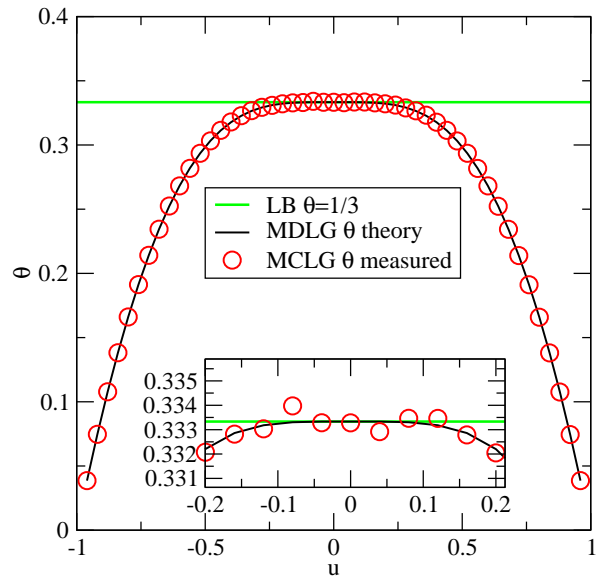}
  \caption{Second velocity moment of Eq. (\ref{eqn:f0_2mom}) divided by the mean density for the MCLG equilibrium distribution for D1Q3. The crucial result here is that in the relevant region $|u|<0.1$ there is excellent agreement between the standard LB result and the MCLG result.} 
  \label{fig:D1Q3tet}
\end{figure}

We show the second moment of Eq. (\ref{eqn:f0_2mom}) for our MCLG equilibrium distribution in Fig. \ref{fig:D1Q3tet}. As predicted, the effective temperature for the MCLG method goes to zero for large absolute velocities, as is unavoidable for a discrete method with a restricted velocity set that has positive definite occupation numbers. For the relevant region where $|u|<0.1$, however there is excellent agreement between the MCLG and LB results, as shown in the inset.

After identifying the global equilibrium distribution for the D1Q3 MCLG, we should now examine the collision operator more closely. 
Eq. (\ref{eqn:1Ddecay}) implies that the collision operator leaves the mass and momentum modes unchanged (as it has to, since those are conserved) and only alters the $\pi$ mode. Let us define its deviation from equilibrium  as $\tilde{\pi}=\pi-\pi^{eq}$.  If we denote $\tilde{\pi}^{(n)}$ as the value of $\tilde{\pi}$ after $n$ collisions we can write the effect of one collision as 
\begin{equation}
  \tilde{\pi}^{(n+1)} = \tilde{\pi}^{(n)} -\frac{\lambda}{2\sqrt{6}}\left(\frac{(\tilde{\pi}^{(n)})^2}{\rho^2}+\frac{\tilde{\pi}^{(n)}}{\rho} 2\sqrt{6} \sqrt{1+3u^2}\right).
  \label{piprime}
\end{equation}
If we interpret this as a non-linear differential equation we can write the analytical solution after $C$ collisions as
\begin{equation}
  \tilde{\pi}^{(C)} = \frac{2\sqrt{6}\rho\sqrt{1+3u^2}\; \tilde{\pi}^{(0)}}{\exp\left(\frac{C\lambda \sqrt{1+3u^2}}{\rho}\right) (2\sqrt{6}\rho \sqrt{1+3u^2}+\tilde{\pi}^{(0)})-\tilde{\pi}^{(0)}}.
  \label{eqn:D1Q3relax}
\end{equation}
For the standard BGK approach in lattice Boltzman we would have expected a pure exponential decay, corresponding the the approximation of neglecting $\tilde{\pi}^2$ in Eq. (\ref{piprime}). The connection between the LG collision operator and the LB BGK collision operator is interesting. The collision MCLG operator is quadratic (and would also involve higher powers if we allowed for multi-particle collisions), but the BGK operator is linear. This apparent mismatch is resolved by writing the collision operator in terms of the moments, and examine only the decay of the non-equilibrium moments. In such a representation, close to local equilibrium, all moments are small, except for the conserved moments, most notably the density. Because of the more complicated dependence of the equilibrium distribution on the momentum, the momentum dependence is more complicated than simply quadratic. So quadratic terms that connect a mode to the local density will dominate the collision operator, and since the density does not change during the collision, the linear relaxation of those moments is recovered.

A BGK collision operator implies that we have the matrix of Eq. (\ref{eqn:Lambda})
\begin{equation}
  \Lambda^{ab}=\left(\begin{array}{ccc}
    0 & 0 & 0\\
    0 & 0 & 0\\
    0 & 0 & 1/\tau^\pi
  \end{array}\right)
\end{equation}
where $\tau^\pi$ is supposed to be a constant independent of the (non-conserved) moments. We can obtain this if we assume $\tilde{\pi}^{(0)}\ll \rho$  and make the usual assumption $u\ll 1$ to get
\begin{equation}
  \tilde{\pi}^{(C)} \approx \exp\left(-\frac{C\lambda}{\rho}\right)\; \tilde{\pi}^{(0)}.
  \label{eqn:D1Q3relax2}
\end{equation}
The relaxation time can then be deduced from the collision term in moment form
\begin{equation}
  \tilde{\pi}^{(C)} = \tilde{\pi}^{(0)}-\frac{1}{\tau^\pi}\tilde{\pi}^{(0)} 
\end{equation}
from which we get
\begin{align}
  \tau^\pi =& \frac{\tilde{\pi}^{(0)}}{\tilde{\pi}^{(0)}-\tilde{\pi}^{(C)}}\\
  \approx& \frac{1}{1-\exp(-C\lambda/\rho)}
\end{align}
which becomes a constant, \textit{i.e.} independent of $\tilde{\pi}^{(0)}$ and $u$, only in the limit mentioned above.
This means that the range or reachable relaxation times are given by $\tau^\pi\in[1,\infty]$. For simple collisions the important over-relaxation regime for $\tau \in [1/2,1]$ that is often utilized in lattice Boltzmann simulations is unavailable for this lattice gas method. An extension of the collision process that includes over-relaxation is possible for high enough densities, but it derivation is outside the scope of this paper.

For systems far from equilibrium or high mach numbers these relaxation times become functions of the state variables in the lattice gas. The dependence of the relaxation time on $u$ violates Galilean invariance, and can in principle be remedied by setting
\begin{equation}
  \lambda= \frac{\tilde{\lambda}}{\sqrt{1+3u^2}}
\end{equation}
for some constant $\tilde{\lambda}$, as can be seen in Eq. (\ref{eqn:D1Q3relax}).

\begin{figure}
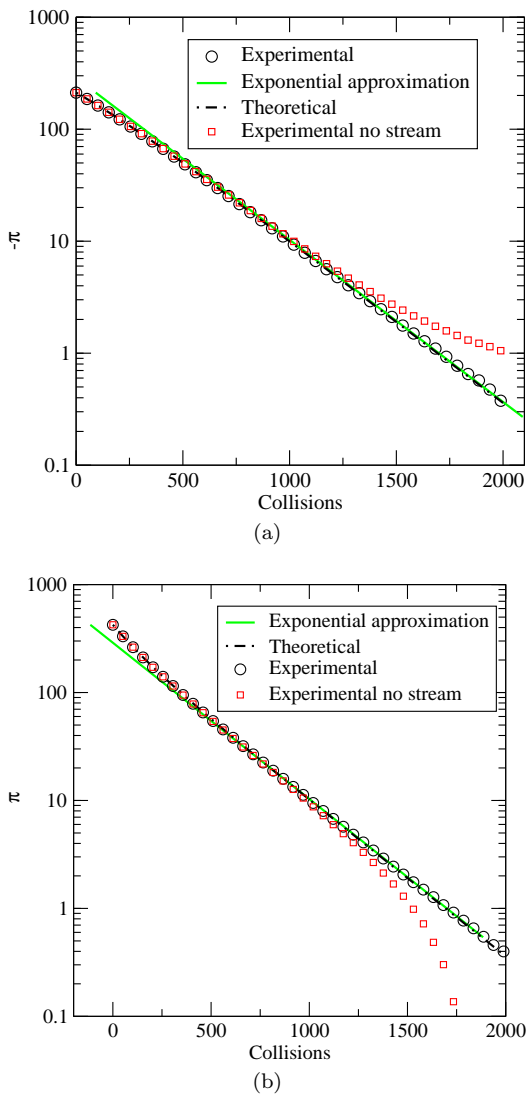

  \centering
  \subfloat[]{\includegraphics[width=0.8\columnwidth]{allrest.eps}}\\
  \subfloat[]{\includegraphics[width=0.8\columnwidth]{movingaway.eps}}
  \caption{Measured and theoretical decay of the second moment of a D1Q3 simulation with 300 particles per cell for two extreme situations. In (a) all particles start as rest-particles, in (b) there are no rest-particles. We observe excellent agreement between the simulation and prediction of Eq. (\ref{eqn:D1Q3relax}) for the relaxation of $\tilde{\pi}$  towards global equilibrium. Also shown is the relaxation towards local equilibrium for a simulation without streaming, which will lead to a negative $\tilde{\pi}$ because of a small difference between the global and local equilibrium, as discussed after Eq. (\ref{eqn:local}). These results are also compared to the linearized theory, leading to an exponential decay of Eq. (\ref{eqn:D1Q3relax2}).}
  \label{fig:decay}
\end{figure}

We test our analytical prediction of the decay of the $\pi$ mode by averaging over 1000 realizations of an initial configuration of $n_i(x,0)=\{100,100,100\}$ on a system with $L_x = 1000$ and $\lambda=1$ and examine the decay of $\langle \pi\rangle$ as a function of the number of collisions. There is a small subtlety in relating the result of the predictions of the Boltzmann averaged lattice gas to actual fluctuating lattice gas simulations. Strictly speaking the analysis we have performed corresponds to relaxations of an ensemble average of systems where the both the local density and the local velocity correspond to the Boltzmann density $\rho$ and Boltzmann velocity $u$ only on average and averaging over the fluctuations is implied. The same is true for the equilibrium distribution: this is an equilibrium average where the realizations averaged over are fluctuating and only the expectation values correspond to the imposed $\rho$ and $u$. 

To compare our Boltzmann theory to simulations we therefore have to consider the average relaxation of $\pi$ in a full simulation, including streaming, of a macroscopically homogeneous system. These results are shown in Figure \ref{fig:decay}, and show excellent agreement between simulation an theory. 

The local equilibrium distribution, i.e. without taking the full Boltzmann ensemble average, has a density dependent second moment. This is easily seen as a single particle with mean velocity zero corresponds to the single state $(n_{-1}=0,n_0=1,n_1=0)$ which has $\pi=-\sqrt{3/2}$. So the local equilibrium distribution in moment representation is
\begin{equation}
  M f_i^0(\rho=1,u=0) = \left(\begin{array}{c}1\\0\\-\sqrt{\frac{1}{2}}\end{array}\right)
\end{equation}
For two particles we would have two states $(n_{-1}=0,n_0=2,n_1=0)$ and $(n_{-1}=1,n_0=0,n_1=1)$ with probability 16/17 and 1/17 respectively according to Eq. (\ref{transprob}). With this we obtain
\begin{equation}
  M f_i^0(\rho=2,u=0) = \left(\begin{array}{c}2\\0\\-\frac{28}{17}\sqrt{\frac{1}{2}}\end{array}\right)
  \label{eqn:local}
\end{equation}
For larger number of particles and different velocities the detailed calculation of the average occupation numbers becomes more complicated, but the existence of a small negative contribution, which does remain of the order of $-1$ and does not increase with the particle number, remains as a part of the local equilibrium distribution.  

Apart from the implementation details, there is an underlying physical reason for this. This second moment is related to a local temperature. It is well known that in a molecular system such a local temperature of a small domain depends on the discretization and at smaller discretization more of the kinetic energy resides in fluctuations of the local velocity and less in the temperature. In the extreme case of a single particle in a discretization cell the temperature would be zero, and all the energy would reside in the momentum of the cell. Therefore the global equilibrium distribution will have a definite momentum and a higher temperature, whereas a local equilibrium will have a fluctuating momentum and a lower temperature. Since the moments here are calculated with respect to the global equilibrium distribution we find that the local equilibrium distributions have small negative deviations of the second moment. This is why the local second moment does not decay towards zero in Figure \ref{fig:decay}, but will become slightly negative instead. In this Figure we see that the full solution of Eq. (\ref{eqn:D1Q3relax}) gives a very good description of the non-linear decay for situations far from equilibrium.

\subsection{D2Q9 results}
The derivation of the D2Q9 collision operator follows the D1Q3 derivation we just showed, but because we have 6 modes that relax, instead of just 1, collision rules become very lengthy to write down here. They are easily derived from the collision classes presented in Figure \ref{fig:D2Q9coll}. We now have nine $\lambda_i$ which make the equivalent of Eq. (\ref{eqn:collmatD1Q3}) again rather lengthy to write. The equivalent of Eq. (\ref{eqn:1Ddecay}) shows a similar structure: many quadratic terms in the $m_i$. Of the $9^3=729$ possible combinations 243 actually occur, each typically associated with several $\lambda_i$. See Supplemental Material at [URL will be inserted by publisher] for a mathematica notebook that contains all the terms in question.

\begin{figure}
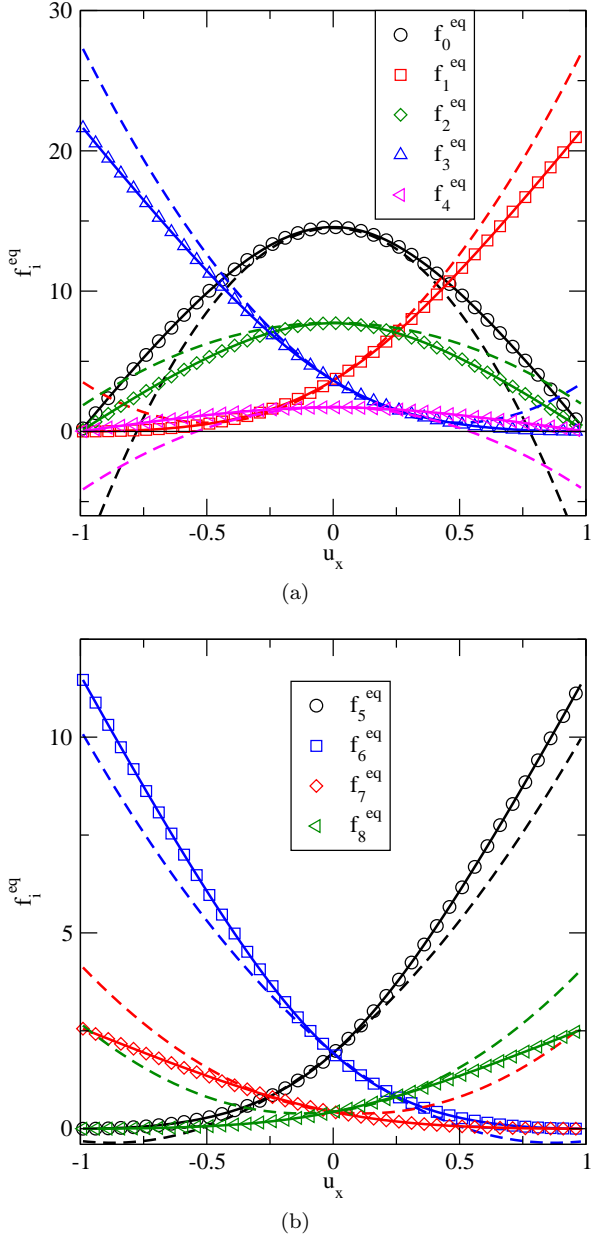

  \subfloat[]{
    \includegraphics[width=0.9\columnwidth,clip=true]{025nivsvx.eps}}\\
  \subfloat[]{
    \includegraphics[width=0.9\columnwidth,clip=true]{025nivsvx58.eps}}
  \caption{Comparison of measured equilibrium distribution for the lattice gas, the theoretical prediction of Eq. (\ref{D2Q9eqdist}), shown as solid lines, and the lattice Boltzmann equilibrium distribution of Eq. (\ref{LBequil}), shown as dotted lines. We show the results as a function of $u_x$ and have set $u_y=0.25$ to visually separate the different averages. For smaller $u_y$ the agreement between MCLG and LB results increases.}
  \label{n0ton4}
\end{figure}

To calculate the equilibrium distribution we need to set $\Omega_i=0$. This initially appeared like a hard problem, and  Mathemactia was unable to solve the resulting equations. However, we noticed that using the product of two one-dimensional equilibrium distribution functions 
\begin{align}
  &f_i^{eq}(\rho,u)\nonumber\\
  =& \rho  \prod_\alpha w_{v_{i\alpha}}\left[1+3 v_{i\alpha}u_\alpha+(3v_{i\alpha}^2-1)(\sqrt{1+3u_\alpha^2}-1)\right]
\label{D2Q9eqdist}
\end{align}
turns out to be the simple solution (no summations over repeated Greek indices is implied here). We did not initially realize this, but at the 2017 DSFD conference in Erlangen AW had a chance to discuss these results with Santosh Ansumali. To both our surprise it turns out that the equilibrium distribution of entropic lattice Boltzmann derived in \cite{ansumali2003minimal} and Eq. (\ref{D2Q9eqdist}) are identical, even if they are written in a very different form. In retrospect this result is not surprising since both methods will minimize the H-function \cite{wagner1998h}
\begin{equation}
  H(\{f_i\}) = \sum_i f_i \log\left(\frac{f_i}{w_i}\right).
  \label{Hfunc}
\end{equation}
To validate that our result for the equilibrium distribution agrees with results of the simulation method we ran a simulation on a lattice with dimensions $L_x=100$, $L_y=10$, $C=10$, $\lambda_1=15/128$, $\lambda_{2-7}=1/4$, $\lambda_8=1/8$, $\lambda_9=18/144$, $\langle\rho\rangle = 36$ and $\langle U_y\rangle = 0.25$. We varied $\langle U_x\rangle$.  We then iterated over 20,000 iterations to ensure we are in equilibrium and then measured for another 10,000 iterations. The results of these simulations are shown in  Fig. \ref{n0ton4}. As expected our values for the $f_i^{eq}$ agree exactly with the measured values. We also compare the quadratic equilibrium distribution of Eq. (\ref{LBequil}) to the values of the MCLG equilibrium distribution (\ref{D2Q9eqdist}) or, equivalently, the entropic LB. They agree around $u_x^{eq}=0$ by design, but for larger velocities the two models begin to diverge.  This is expected because as the mean velocity increases, the $f_i^{eq}$ of Eq. (\ref{D2Q9eqdist})  eventually become negative, which the $n_i$ cannot do. Note that we have picked a very large $U_x=0.25$ to emphasize the differences between the entropic and polynomial equilibrium distributions.

As an example we show here the relaxation of the stress mode, which is responsible for recovering the shear viscosity. Similar to our analysis for the D1Q3 case we examine the effect of the collision on the deviation of the distribution from the local equilibrium distribution. We write the non-equilibrium part of our moments as
\begin{align}
  dM^a &= \sum_i m_i^a (f_i - f_i^{eq})\\
     &= \left(\begin{array}{c}
    0\\
    0\\
    0\\
    \sigma_--\rho(A_x-A_y)\\
    \sigma_{xy}-\rho j_x j_y\\
    \sigma_+-\rho(A_x+A_y-2)\\
    q_x-\sqrt{2}j_x(A_y-1)\\
    q_y-\sqrt{2}j_y(A_x-1)\\
    t-2\rho(A_x-1)(A_y-1)
  \end{array}\right)
  \label{eqn:dMD2Q9moments}
\end{align}
where we introduced
\begin{align}
  A_x &= \sqrt{1+\frac{j_x^2}{\rho^2}};
  &A_y &= \sqrt{1+\frac{j_y^2}{\rho^2}}
\end{align}
so that $dM^a=0$ in equilibrium. We write the effect of a single collision on these non-equilibrium moments as
\begin{align}
   dM^a(n) = \Omega^{a,(n)}=\sum_i m_i^a \Omega_i^{(n)}
\end{align}
and we replace the appearances of $f_i$ in the collision term with
\begin{equation}
  f_i = w_i \sum_a m_i^a (dM^a+M^{a,eq})
\end{equation}
The conservation of mass and momentum immediately imply that $\Omega^{0(n)}=\Omega^{1(n)}=\Omega^{2(n)}=0$. For the collision term related to the stress moment $\sigma_{xy}$ we get the rather lengthy
\begin{align}
 & \sigma_{xy}^{(n+1)}=\sigma_{xy}^{(n)}
  +\frac{1}{9\rho} [\left(-8 \left(-2+A_x\right) \left(-2+A_y\right) \rho \right) \lambda _ 1\nonumber\\
  & +\left(-\left(-1+2 A_x\right) \left(-1+2 A_y\right) \rho \right) \lambda _ 3\nonumber\\
  &+(-2 +4 A_x +4 A_y -8 A_x A_y) \rho  \lambda _ 4 \nonumber\\
  &+(-2 +4 A_x +4 A_y -8 A_x A_y) \rho  \lambda _ 6 \nonumber\\
  &+( 8-10 A_x-10 A_y +8 A_x A_y) \rho  \lambda _ 7 \nonumber\\
  &+(16-20 A_x-20 A_y+16 A_x A_y) \rho  \lambda _ 8\nonumber\\
&   +(8 \lambda_1
    -2 \lambda_3
    -4 \lambda _ 4
    -4 \lambda _ 6
    -2 \lambda _ 7
    -4 \lambda _ 8)\sigma _+^{(n)}\nonumber\\
    &+(-4  \lambda_1
    -2 \lambda_3
    -4  \lambda _ 4
    -4  \lambda_6
    +4  \lambda _ 7
    +8  \lambda _ 8)t^{(n)}]\sigma_{xy}^{(n)}\nonumber\\
  & + \frac{2j_xj_y}{9\rho^2}(4 \lambda _ 1-\lambda _ 3-2 \lambda _ 4-2 \lambda _ 6- \lambda _ 7-2 \lambda _ 8)\sigma_+^{(n)}\nonumber\\
  & + \frac{2j_xj_y}{9\rho^2}(2 \lambda _ 1+\lambda _ 3+2 \lambda _ 4+2 \lambda _ 6-2\lambda _ 7-4 \lambda _ 8)t^{(n)}\nonumber\\
  & + \frac{2     }{9\rho    }(2 \lambda _ 1+\lambda _ 3+2 \lambda _ 4+2 \lambda _ 6-2\lambda _ 7-4 \lambda _ 8)q_x^{(n)} q_y^{(n)}\nonumber\\  &+[4(A_x-2)\lambda_1+(2A_x-1)\lambda_3+2(2A_x-1)\lambda_4\nonumber\\&+2(2A_x-1)\lambda_6-(4A_x-5)\lambda_7-(8A_x-10)\lambda_8]\frac{\sqrt{2}j_y q_x^{(n)}}{9\rho}\nonumber\\ &+[4(A_y-2)\lambda_1+(2A_y-1)\lambda_3+2(2A_y-1)\lambda_4\nonumber\\&+2(2A_y-1)\lambda_6-(4A_y-5)\lambda_7-(8A_y-10)\lambda_8]\frac{\sqrt{2}j_x q_y^{(n)}}{9\rho}\nonumber
\end{align}
This expression shows the coupling of the various modes. Unlike in the one-dimensional case we were unable to find an analytical solution for this non-linear relaxation. However, if we restrict our attention to moments that are close to local equilibrium, then the only terms that are not small are the conserved quantities. Here there are three conserved quantities, $\rho$, $j_x$, and $j_y$. And since $j_x=\rho u_x$ and $u_x<0.1$ for typical applications we can make the further assumption $j_x^2\ll \rho^2$. This means that the leading order terms are contained in the first 6 lines. The next order, small only because $|j|\ll \rho$, are the tems in the last four lines. The terms in lines 9 and 10 are smaller by one factor of $j_x/\rho\ll 1$.  The remaining terms are quadratic in the non-equilibrium moments (remember those are zero in equilibrium) and can be neglected for systems close to local equilibrium. To leading order, neglecting $|j|/\rho$ terms, we also have $A_x=A_y=1$. The leading order for the relaxation of the $\sigma_{xy}$ mode is then given by
\begin{align}
   \sigma_{xy}^{(n+1)}=\sigma_{xy}^{(n)}-\frac{1}{9\rho}[8 \lambda _ 1+\lambda _ 3+2 (\lambda _ 4+\lambda _ 6+2 \lambda _ 7+4 \lambda _ 8)]\sigma_{xy}
\end{align}
Analogously to the one-dimensional case we can derive the effective relaxation time as
\begin{equation}
  \tau^{\sigma_{xy}} = \frac{1}{1-\exp\left(-\frac{C\left(8\lambda _ 1+\lambda _ 3+2\left(\lambda _ 4+\lambda _ 6+2 \lambda _ 7+4 \lambda _ 8\right)\right)}{9\rho}\right)}
  \label{tausigma}
\end{equation}
Applying this approximation for all of the modes we obtain
\begin{widetext}
\begin{align}
\Omega^{a,(n)}  &= \frac{2}{\rho}\left(\begin{array}{c}
 0 \\
 0 \\
 0 \\
  \sigma_-^{(n)} \left(-\frac{2}{9} \lambda _2-\frac{1}{36} \lambda _4-\frac{4}{9} \lambda _5-\frac{2}{9} \lambda _7-\frac{1}{36} \lambda _9\right)\\
 \sigma_{xy}^{(n)} \left(-\frac{4}{9} \lambda _1-\frac{1}{18} \lambda _3-\frac{1}{9} \lambda _4-\frac{1}{9} \lambda _6-\frac{2}{9} \lambda _7-\frac{4}{9} \lambda _8\right) \\
 -\frac{2}{9}  \left((\sigma_+^{(n})-t^{(n)}) \lambda _2+(\frac{1}{4} \sigma_+^{(n)}+\frac{1}{8} t^{(n)}) \lambda _3+(\frac{1}{8} \sigma_+^{(n)} +\frac{1}{4} t^{(n)}) \lambda _4+ (\sigma_+^{(n)} +\frac{1}{2} t^{(n)}) \lambda _7+(\frac{1}{8} \sigma_+^{(n)} +\frac{1}{4} t^{(n)}) \lambda _9\right) \\
  q_x^{(n)} \left(-\frac{2}{3} \lambda _1-\frac{1}{3} \lambda _7-\frac{1}{3} \lambda _8-\frac{1}{12} \lambda _9\right) \\
  q_y^{(n)} \left(-\frac{2}{3} \lambda _1-\frac{1}{3} \lambda _7-\frac{1}{3} \lambda _8-\frac{1}{12} \lambda _9\right) \\
 - \left(t \lambda _1+(\frac{2}{9} t^{(n)}-\frac{2}{9} \sigma_+^{(n)}) \lambda _2+(\frac{1}{36} \sigma_+^{(n)}+\frac{1}{72} t) \lambda _3+(\frac{1}{18} \sigma_++\frac{1}{9} t) \lambda _4+(\frac{1}{9} \sigma_++\frac{1}{18} t) \lambda _7+(\frac{1}{18} \sigma_+^{(n)}+\frac{1}{9} t^{(n)}) \lambda _9\right) 
\end{array}\right)
\label{D2Q9MOmega}
\end{align}
\end{widetext}
We see that the relaxation of all moments decouples, except for the $\sigma_+$ and $t$ moments. To recover a diagonal collision matrix in moment space $\Lambda^{ab}$ is equivalent to decoupling here. The 8th moment couples to $\sigma_+$ with a the same coefficient that the 5th moment couples to the eighths. We can set this coefficient to zero by demanding:
\begin{equation} \frac{2}{9}\lambda_2-\frac{1}{36}\lambda_3-\frac{1}{18}\lambda_4-\frac{1}{9}\lambda_7-\frac{1}{18}\lambda_9=0
\label{decouple}
\end{equation}
which then decouples the modes and recovers the diagonal collision matrix for D2Q9. 
Additionally, to recover an isotropic stress tensor, we demand that the $\sigma_{xy}$ and $\sigma_-$ be relaxed at the same rate:
\begin{align}  
  &\left(-\frac{2}{9} \lambda _2-\frac{1}{36} \lambda _4-\frac{4}{9} \lambda _5-\frac{1}{36} \lambda _9\right)\nonumber\\
  =&\left(-\frac{4}{9} \lambda _1-\frac{1}{18} \lambda _3-\frac{1}{9} \lambda _4-\frac{1}{9} \lambda _6-\frac{4}{9} \lambda _8\right).
  \label{rotational}
\end{align}
The moments $q_x$ and $q_y$ are automatically relaxed at the same rate.

To derive the Boltzmann average of this MCLG method of Eq. (\ref{MCLB}) we first need to write the effect of $C$ collisions for Eq. (\ref{D2Q9MOmega}). Once we ensured that the matrix terms decouple by imposing Eq. (\ref{decouple}) we obtain for the collision matrix
\begin{equation}
  \Lambda^{ab} = \frac{1}{\tau^a} \delta^{ab}
 \label{D2Q9Lambda}
\end{equation}
where the relaxation times are given by (\ref{tausigma}) for the $\sigma_{xy}$ moment and equivalent expressions for the relaxation of the other modes. 
This means we have derived a multi-relaxation-time BGK collision operator (with the assumption that we are close enough to equilibrium that the density mode is much larger than the non-conserved moments and that $\rho\gg |j|$) from the underlying lattice gas collision rules. We believe that this is the first such derivation. Earlier introductions of a BGK collision operator \cite{Higuera1989,qian1992lattice} were unable to derive this from a lattice gas, and instead did so as an ad-hoc assumption. The derivation is quite general: in moment space the dominant terms are those which are proportional to the density, and these leading order terms will contain only one power of one non-equilibrium moment, meaning that the collision operator can be written as a linear operator, \textit{i.e.} a matrix collision applied to the non-equilibrium part of the particle distribution $f_i-f_i^{eq}$.

The Boltzmann average of the Monte Carlo Lattice Gas algorithm, with appropriate approximations discussed above, can now be written as
\begin{equation}
  f_i(x+v_i,t+1) = f_i(x,t)+\sum_j \Lambda_{ij} [f_j^{eq}(\rho,u)-f_j(x,t)]
\end{equation}
where the collision matrix is given by Eq. (\ref{D2Q9Lambda}) and the local equilibrium by Eq. (\ref{D2Q9eqdist}).

\begin{figure}
  \includegraphics[width=\columnwidth]{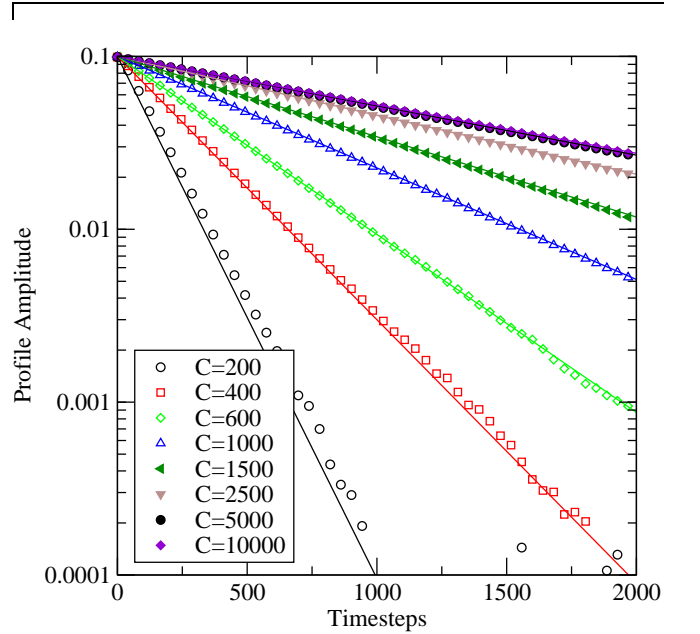}
  \caption{Comparison of predicted (solid lines) and measured (symbols) decay of a shear wave with initial amplitude of $A = 0.1$. We observe that the decay rate corresponds closely to the rate predicted by Eqn. (\ref{Eqn:Decay}) and Eq. (\ref{Eqn:nu}), except for the smallest number of collisions coresponding to a relaxation time of $\tau=2.69$.}
  \label{Fig:Decay}
\end{figure}

Since this is exactly the Entropic lattice Boltzmann method (without over-relaxation for the moment) the standard derivations show that we recover the continuity 
\begin{equation}
  \partial_t \rho + \nabla_\alpha (\rho u_\alpha) =0
\end{equation}
and Navier-Stokes equation
\begin{align}
  &\partial_t(\rho u_\alpha)+ \nabla_\beta(\rho u_\alpha u_\beta) \nonumber\\&= \nabla_\alpha p +\nabla_\beta[\rho \nu (\nabla_\alpha u_\beta+\nabla_\beta u_\alpha-\frac{2}{3}\nabla_\gamma u_\gamma \delta_{\alpha\beta})]
\end{align}
with
\begin{equation}
  \nu = \left(\tau^{\sigma_{xy}}-\frac{1}{2}\right)\theta
  \label{Eqn:nu}
\end{equation}
To validate our prediction for the shear viscosity we set up a sinusoidal shear profile
\begin{align}
  \rho(x,y,0) &= \rho^0,\nonumber\\
  u_x(x,y,0) &= A \sin\left(\frac{2\pi y}{L_y}\right),\\
  u_y(x,y,0) &= 0\nonumber
\end{align}
for which the Navier-Stokes equation has the analytical solution
\begin{align}
  \rho(x,y,t) &= \rho^0,\nonumber\\
  u_x(x,y,t) &= A \exp\left(-\frac{4\pi^2\nu t}{L_y^2} \right)\sin\left(\frac{2\pi y}{L_y}\right),\label{Eqn:Decay}\\
  u_y(x,y,t) &= 0.\nonumber
\end{align}
To simulate this system with the MCLG method we set up the lattice densities as 500 particles per lattice site, $L_x=100, L_y=101$, $\lambda$ values the same as those given below Eq. (\ref{Hfunc}). To initialize the profile we set only the rest density $n_0$ and either $n_1$ or $n_2$ different from zero such that $\sum_i n_i=500$ and $\sum_i n_i v_{ix}=500\times 0.1 \sin(2\pi y/L_y)$, where the last equality is rounded to the nearest integer value. The densities are then equilibrated by performing 10,000 collisions on each lattice site (without streaming) to generate the initial configuration for the start of the simulation.
The results of our MCLG simulations show a fluctuating version of the decay of the sinusoidal  profile of such a shear wave. We average the results by averaging over the 100 $x$ values. To reduce fluctuations further we averaged this over $15$ separate realizations of this simulation. The simulation results show clean sinusoidal profiles with amplitudes that decay exponentially as a function of time. We obtain the amplitude from Eq. (\ref{eqn:A}), given in an appendix. The results are shown in Figure \ref{Fig:Decay}, where we see that there is excellent agreement between the decay of the shear wave predicted by theory and the simulations, thus confirming our predictions for the shear viscosity in the MCLG method.

\section{Fluctuating properties of the MCLG}
So far we have focused on the Boltzmann average of our lattice gas. The original reason we were interested in the lattice gas were its fluctuating properties. The fluctuations in an ideal gas have been discussed by Landau \cite[\S 114]{landau1969statistical}, where he shows that for a classical Boltzmann gas the number density in sub-volumes are Poisson distributed. The argument here is trivially extended to lattice gases showing that each density $n_i$ should be Poisson distributed:
\begin{equation}
P(n_i)=\frac{exp(-f_i^{eq})(f_i^{eq})^{n_i}}{n_i!}.
\label{poisson}
\end{equation}
As a consequence we obtain
\begin{equation}
  \langle n_i(x,t)n_j(y,t)\rangle = f_i^{eq} f_j^{eq}+f_i^{eq} \delta_{ij}\delta_{xy}.
  \label{eqn:Poisson2}
\end{equation}
In the lattice Boltzmann context trying to recover a Poisson distribution is problematic, since the densities $f_i$ are continuous, and there is no generally accepted extension of the Poisson statistic to continuous variables.
Instead of trying to impose a Poisson distribution,  Eq. (\ref{eqn:Poisson2}) is used as the starting point for the derivation of fluctuating lattice Boltzmann methods. Using this second moment one finds that the lattice Boltzmann densities approximately represent a Poisson distribution, as seen in recent papers by Kaehler et al. \cite{kaehler2013fluctuating} as well as by Wagner et al. \cite{wagner2016fluctuating}. However, these solutions are never exact and it is important to be aware of where the assumption of Poisson distributed $f_i$, which is used as an input to derive these methods, breaks down in the practical implementation. 

\begin{figure}
  \centering
  \includegraphics[clip=true,width=\columnwidth]{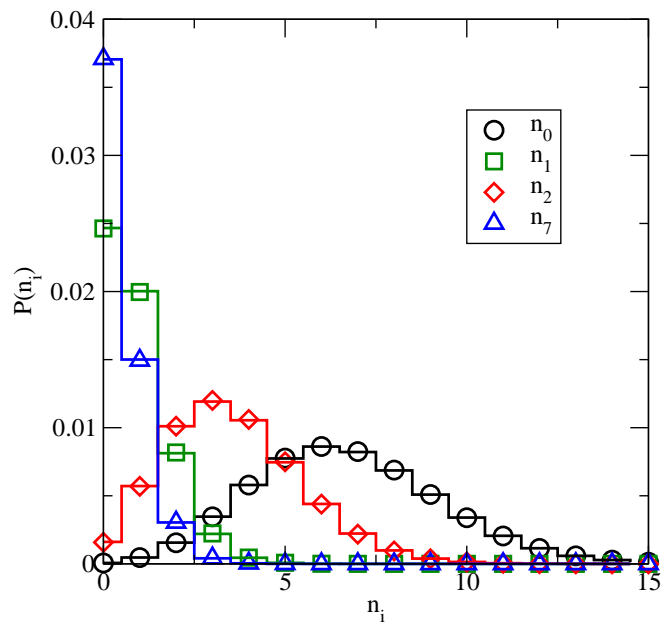}
  \caption{$n_0$, $n_1$, $n_2$, $n_7$, plotted with respect to local density at $U_x=-0.25$, $U_y=0.25$ and $\rho=18$. Symbols represent the measured values while the solid lines are the theoretical distribution.}
  \label{fig:Poisson}
\end{figure}

\begin{table*}
  \begin{tabular}{r|rrrrrrrrr}
    &$n_0$&$n_1$&$n_2$&$n_3$&$n_4$&$n_5$&$n_6$&$n_7$&$n_8$\\\hline
$n_0$& 0.998863& -0.000028& -0.000482& -0.000170& -0.000343& 0.000001& -0.000114& -0.000082& 0.000010\\
$n_1$&-0.000028&  0.999171& 0.000001& 0.000143& -0.000072& -0.000127& -0.000047& 0.000013& 0.000057\\
$n_2$&-0.000482& 0.000001&  0.999809& -0.000112& 0.000177& -0.000515& -0.000110& -0.000022& 0.000482\\
$n_3$&-0.000170& 0.000143& -0.000112& 1.000230& -0.000079& 0.000014& 0.000008& -0.000048& -0.000067\\
$n_4$&-0.000343& -0.000072& 0.000177& -0.000079& 0.999648& 0.000492& -0.000023& -0.000103& -0.000480\\
$n_5$&0.000001& -0.000127& -0.000515& 0.000014& 0.000492& -0.999567& -0.000065& -0.000046& 0.000593\\
$n_6$&-0.000114& -0.000047& -0.000110& 0.000008& -0.000023& -0.000065&  0.999801& -0.000085& 0.000086\\
$n_7$&-0.000082& 0.000013& -0.000022& -0.000048& -0.000103& -0.000046& -0.000022&  0.999687& -0.000117\\
$n_8$&0.000010& 0.000057& 0.000482& -0.000067& -0.000480& 0.000593& 0.000086& -0.000117& 0.999215
  \end{tabular}
  \caption{Representation of the correlator of Eq. (\ref{eqn:corr}) for pairs of $\langle n_i n_j\rangle$ for the worst statistical outlier for $u_x=-0.68$ of Figure \ref{fig:error}.}
  \label{table1}
\end{table*}

\begin{figure}
  \centering
  \includegraphics[width=0.9\columnwidth]{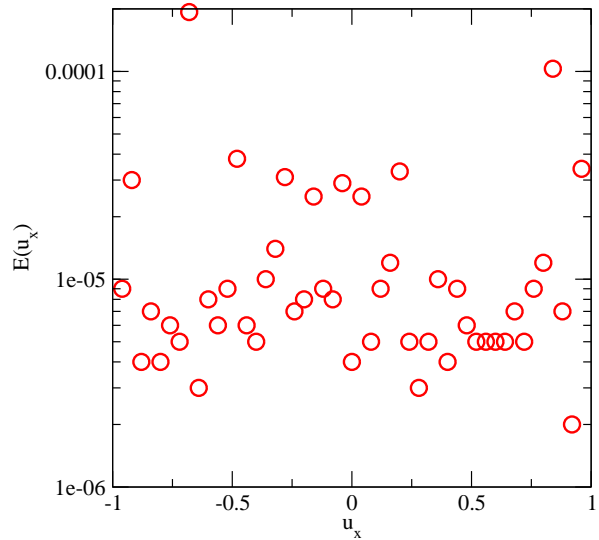}
  \caption{The deviation of the expected Poisson distribution statistic from the measured correlations from Eq. (\ref{eqn:error}) for different velocities $u$. There is no discernible pattern, and the error continues to decrease with increased averaging. }
  \label{fig:error}
\end{figure}

It is important to note that this result is independent on the transport parameters, or in our case the number of collisions $C$, the various acceptance prefactors $\lambda_i$, the density or an imposed fluid velocity. We therefore tested the prediction of Poisson distributed occupation numbers $n_i$ in the MCLG method for a large variety of densities and imposed velocities and throughout found excellent agreement with the prediction from Eq. (\ref{poisson}). An example is shown in Figure \ref{fig:Poisson}. Similarly we can test the independence of the $n_i$ by looking at
\begin{equation}
  \frac{ <n_i n_j>-f_i^{eq}f_j^{eq}}{\sqrt{f_i^{eq} f_j^{eq}}}\stackrel{?}{=}\delta_{ij}.
  \label{eqn:corr}
\end{equation}
Again we found excellent agreement for all densities and velocities that we tested. For the results shown here we used the same values for $\lambda_i$ as above, we initialized the lattice with a density of 360 and an initial velocity $u_x$ with $u_y=0$. We performed $C=10$ collisions per site per iteration.  We then discarded the first 150,000 iterations to ensure we are looking at an equilibrium systems. We averaged over the next 500,000 steps on a lattice with $L_x=100$, $L_y=10$. We show the worst case result with respect to the results of Fig. \ref{fig:error} in Table \ref{table1}. We see that we have excellent agreement with our prediction of Eq. (\ref{eqn:corr}).

To examine the agreement for different velocities without printing many of these tables we look at a measure of the deviation of the correlation matrix from the expected value
\begin{equation}
  E(u_x,u_y)=  \sum_{i,j}  \left(\frac{ <n_i n_j>-f_i^{eq}f_j^{eq}}{\sqrt{f_i^{eq} f_j^{eq}}}-\delta_{ij}\right)^2.
  \label{eqn:error}
\end{equation}
A graph of this error function as a function of the mean velocity is shown in Figure \ref{fig:error}. We see that there is no apparent systematic variation of this error function with $u$ even for extreme values for $u$ and with increasing averaging the error continues to diminish.
This suggests that we have found a reference lattice gas implementation that exactly fulfills the predictions for a fluctuating ideal gas, a feat that has not been fully achieved to date for fluctuating lattice Boltzmann methods \cite{kaehler2013fluctuating}.

As a consequence of the correctly Poisson distributed fluctuations of the $n_i$ follows that all equal time corellators of composite quantities like the density or the momentum are correct and obey their respective Poisson or Skelam distributions.

\section{Conclusions}
In this article we have introduced an integer lattice gas method which employs a Monte Carlo collision operator. We show that the Boltzmann limit of this lattice gas recovers the entropic lattice Boltzmann method, which agrees with the usual lattice Boltzmann distribution with a polynomial distribution function for moderate velocities $|u|<0.1$. Remarkably we were able to derive a BGK collision operator directly from the lattice gas collision rules. Previous the BGK collision operator had been postulated only, but not derived. The equilibrium distribution recovered here is identical with that of entropic lattice Boltzmann \cite{ansumali2003minimal} and Boltzmann average of the lattice gas obeys the standard H-theorem.

Remarkably the fluctuating properties of this MCLG recover Poisson statistics exactly in a fully Galilean invariant manner. This suggests that an optimized version of this MCLG method may become a promising contender for the simulation of fluctuating fluids. Extending this result to non-ideal fluids will require a better understanding of fluctuations in these systems. We will examine what fluctuations for non-ideal systems with the help of the Molecular Dynamics lattice gas (MDLG) introduced by Parsa et al. \cite{parsa2017lattice}. Developing an implementation that can compete with lattice Boltzmann approaches with regard to speed will require an implementation of a lattice gas collision operator that can perform multiple collisions at one time and a way to introducing over-relaxation that has been the way of lowering transport coefficients below the value obtained for full relaxation to local equilibrium. This is the subject of current research, and we anticipate to present solutions to these questions in subsequent publications.

\appendix
\section{Amplitude of sine-wave decay}
We can look at the problem of finding the amplitude $A(t)$ that corresponds to the best fit of a profile $u_x^{th}=A(t)\sin(2\pi y/L_y)$ to some data $U_x(t,y)$. The mean-square deviation $E(A)$ would be given by
\begin{align}
  E(A) = \sum_y [U_x(y)-u_x(y)]^2
\end{align}
and the requirement of this being minimal gives us
\begin{align}
  0&=\frac{d E(A)}{dA}\nonumber\\
  &=\frac{d}{dA}\sum_y \left[A^2\sin^2\left(\frac{2\pi y}{L_y}\right)+A\sin^2\left(\frac{2\pi y}{Ly}\right)U_x(y)+U_x^2(y)\right]\nonumber\\
  &=\sum_y \left[2A\sin^2\left(\frac{2\pi y}{L_y}\right)+\sin\left(\frac{2\pi y}{L_y}\right) U_x(y)\right]
\end{align}
so we get
\begin{equation}
  A(t) = \frac{\sum_y\sin\left(\frac{2\pi y}{L_y}\right) U_x(y)}{\sum_y\sin^2\left(\frac{2\pi y}{L_y}\right)}.
\label{eqn:A}
\end{equation}
We can look at $E(A)/A^2$ as a measure for the quality of the fit.

%\begin{acknowledgments}
%\end{acknowledgments}

\bibliography{AW,MCLG,IntegerLG}

\end{document}